# Practical Recovery Solution for Information Loss in Real-Time Network Environment

Hengky Susanto and ByungGuk Kim
Department of Computer Science
University of Massachusetts at Lowell
{hsusanto, kim}@cs.uml.edu

**Abstract** — Feedback mechanism based algorithms are frequently used to solve network optimization problems. These schemes involve users and network exchanging information (e.g. requests for bandwidth allocation and pricing) to achieve convergence towards an optimal solution. However, in the implementation, these algorithms do not guarantee that messages will be delivered to the destination when network congestion occurs. This in turn often results in packet drops, which may cause information loss, and this condition may lead to algorithm failing to converge. To prevent this failure, we propose least square (LS) estimation algorithm to recover the missing information when packets are dropped from the network. The simulation results involving several scenarios demonstrate that LS estimation can provide the convergence for feedback mechanism based algorithm.

*Index Terms* — networks, congestion control, resource management, optimization methods, QoS.[1]

## I. INTRODUCTION

Dynamic pricing is a widely adopted means to overcome network congestion. The relationship between pricing and traffic management is often formulated into Network Utility Maximization (NUM) framework [1,2,3,4, 5,12,15,22]. The common approach to solve NUM problems typically involves feedback mechanism that converges to an optimal solution. That is, the network adjusts the price to control the level of congestion, and users adapt their transmission rate according to the price decided by the network [3,4,5]. The feedback mechanism can be accomplished by the network sending messages containing network price; users then acknowledge the price by responding to it with a message that contains their new demand for bandwidth. This approach is applicable for cloud computing, where users are required to pay according to their network usage.

However, one major shortcoming of this approach is the reliance on explicit communication of both price notification and response message for bandwidth demand. During network congestion, packets that carry the price notification and response message are susceptible to packet loss or being dropped from network. A proper response message is particularly important to determine the appropriate price update interval. It is because, without response message, the exact time interval to broadcast price update is difficult to estimate. In our previous work [13], we have shown that premature price update leads to algorithm oscillation and delayed update leads to slower convergence.

To address this problem in this paper, we propose a solution based on *Least Square (LS) estimation* [20] algorithm to estimate the appropriate time interval for the next update. This technique estimates based on the history of the aggregated inputs of network price and update interval. The strength of LS estimation is that network does not need to see the overall picture of what is happening in the network. Instead, it leverages on information accumulated over time to make an estimation. In this paper, LS estimation algorithm is incorporated into feedback mechanism-based solution to resolve situations when there is information lost. LS estimation can be resolved in linear time and requires only a small memory space, yet it effectively recovers information lost in most situations.

There has been much research done on feedback mechanism such as: Subgradient based algorithms that are often employed to resolve congestion or bandwidth provisioning problem in NUM [2,3,14,15,19], Explicit Congestion Notification (ECN) [6][7], and different variations of feedback mechanism schemes on congestion control [8,10,11]. However, time interval of notification update in feedback mechanism based solutions is not addressed in these literatures. It is because in NUM the information on price and transmission rate is assumed to be available instantaneously. Moreover, the proposed solutions in [6,7,8,10,11] are TCP/IP based, where the network relies on users to manage their own transmission rate through some control methodology like the congestion window [21], thus precise update interval is not necessary in these approaches. Authors of [18] discuss this issue and claim that when the estimator of gradient based algorithm is biased, the solution can still converge to a contraction region around the optimal point, even without complete information. In other words,

---

depending on the choice of step size, the algorithm still converges even with missing information. However, determining the appropriate estimator relies on the assumption that the interval time for update is constant, which may not be the case when network traffic fluctuates.

We address this gap in interval notification update by providing a technique to estimate an appropriate interval, especially when there is a lack of information caused by packet loss. We begin our proposal with problem formulation in section II, where we introduce NUM, subgradient based algorithm, and message exchange mechanism. Following this, we present our major contributions: the mathematical model for interval update for price notification and the design for LS estimation algorithm in section III and IV respectively. The simulation results are presented and discussed in section V, followed by concluding remarks in section VI.

## II. PROBLEM FORMULATION

We begin with a discussion on NUM to support real-time traffic with delay QoS. Consider a network with a set of links $L$, and a set of link capacities $C$ over the links. Given a utility function $U_s(x_s)$ of user $s$ with an allocated bandwidth of $x_s$, the NUM formulation becomes

$$maximize \sum_{s \in S} U_s(x_s) \quad (P)$$

$$s.t. \ Ax \leq C$$

$$over \ x \geq \bar{0}$$

where $S$ and $A$ denote sets of users and routing paths, respectively, and $\bar{0}$ is a vector of zeros. A route $r$ consists of a series of links $l$ such that $A_{lr} = 1$ if $l \in r$ and $A_{lr} = 0$, otherwise. The user utility function is defined as follows.

$$U_s(x_s) = \frac{1}{(1 + e^{-x_s})} \quad (1)$$

The NUM formulation is solved by the Lagrangian method. Typically, a dual problem to the primal problem of (P) is constructed as follows.

$$L(x, \lambda) = \sum_{s \in S} U_s(x_s) - \lambda^T (C - Ax),$$

$$= \sum_{s \in S} U_s(x_s) - \sum_{s \in S} \lambda_s x_s + \sum_{l \in L} \lambda_l C_l,$$

where the Lagrangian multipliers $\lambda_s$ are interpreted as the link costs and

$$\lambda_s = \sum_{l \in r_s} \lambda_l.$$

The dual problem of (P) is then defined as

$$\min D(\lambda) \quad (D)$$
$$s.t \ \lambda \geq \bar{0},$$

where the dual function

$$D(\lambda) = \max_{\bar{0} \leq x \leq x^{max}} L(x, \lambda).$$

The transmission rate $x_s(\lambda_s)$ of user $s$ at link cost $\lambda_s$ can be computed in a distributed manner by

$$x_s(\lambda_s) = \arg\max_{0 \leq x \leq x^{max}} (U_s(x_s)), \quad (2)$$

A subgradient projection method is used in [3], where the network on each link $l$ updates $\lambda_s$ on that link, resulting in an iterative solution given by

$$\lambda^{(t+1)} = \left[ \lambda^{(t)} - \sigma^t \left( C - A \, x(\lambda^{(t)}) \right) \right]^+, \quad (3)$$

where $x(\lambda^{(t)})$ is the solution of (3) and $C - A \, x(\lambda^{(t)})$ is a subgradient of $D(\lambda)$ at link price $\lambda = \lambda^{(t)}$ and $x(\lambda^{(t)})$ denotes the rate allocation at $\lambda^{(t)}$, for $\lambda^{(t)} \geq \lambda^{min}$, where $\lambda^{min}$ can be interpreted as the network's operation cost [16]. Also, $x(\lambda^{(t)})$ denotes the rate allocation at $\lambda^{(t)}$ and $\sigma^{(t)}$ denotes the step size to control the tradeoff between a convergence guarantee and the convergence speed, such that

$$\sigma_l^{(t)} \to 0, \text{as } t \to \infty \text{ and } \sum_{t=1}^{\infty} \sigma_l^{(t)} = \infty.$$

The feedback loop in the pair of equation (2) and (3) allows users to adjust their transmission rate according to the price, and for the network to control the amount of traffic flow by adjusting the price until $\lambda_l^{(t)}$ converges to a solution.

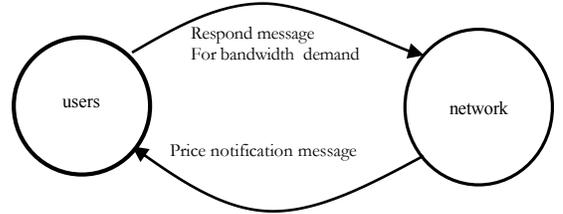

Fig. 1. Message Passing Protocol.

In practice, message exchanges between users and network in subgradient based algorithm can be described as follows. When congestion is detected at link $l$, network (router at link $l$) broadcasts the price notification to users whose flow traverse through it. Then, users response with an response message upon receiving the price notification. The network observes whether traffic condition improves after user's response messages from the entire users on $l$ are received. The relationship between users and the network is illustrated in figure 1. In practice, subgradient based algorithm can be implemented by leveraging an existing protocol, Internet Control Message Protocol (ICMP) [21]. The ICMP packet can be adopted for price notification packet scheme by taking advantage of the unused reserved type in ICMP header, shown in figure 2. Similarly, this concept can be considered for users' response for bandwidth demand message, as depicted in figure 3.

| Type = 1 | Code =0 | Header check sum |
|---|---|---|
| Identifier | | Sequence Number |
| Time Stamp | | *Link Price* |

Fig. 2. ICMP header for price notification.

| Type = 1 | Code =1 | Header check sum |
|---|---|---|
| Identifier | | Sequence Number |
| Time Stamp | | *Bandwidth Request* |

Fig. 3. ICMP header of the user's response message.

However, ICMP protocol does not guarantee packets will reach the destination, especially during the congestion. As a result, network may not receive response messages from users and not be able to determine the appropriate time to broadcast the next price notification. For this reason, we first determine the duration required for packet that carries the price notification and user response to travel from one end to the other end by utilizing M/M/1 queuing model [9].

## III. INTERVAL UPDATE

Information delay is unavoidable in any system that employ feedback mechanism. Thus, we employ Little theorem [9] to model the delay for the price notifications and user's response messages to reach their destination. First of all, the average delay $d_s$ between the destination and where the notification is originated can be estimated as follows.

$$d_s = \sum_{l \in r(s)} \left( \frac{\rho_l}{(x_l - x_l^{min})} + \frac{1}{x_s^{min}} \right), \quad (4)$$

where $x_l^{min}$ is the minimum time required to process notification packet, $x_l$ is amount of bandwidth allocated on link $l$, and the notification along path $r(s)$, which is associated to user $s$. Here, bandwidth $x_l^{min}$ can be interpreted as user minimum requirement for bandwidth allocation. The processing ratio $\rho_l$ is defined as

$$\rho_l = \frac{x_l^{min}}{x_l}.$$

So, the time required $d_s^{info}$ for the price notification packet to reach is destination, the user $s$, can be estimated by

$$d_s^{info} = d_s + \sum_{l \in r(s)} d_l^{propa}$$

where $d_l^{propa}$ denotes the negligible propagation delay at link $l$. Thus, with this estimation, the longest delay $d_l^{max}$ from $l$ to every user that uses link $l$ is

$$d_l^{max} = max\left(d_s^{info} | s \in S(l)\right), \quad (5)$$

where $S(l)$ is a set of users who use link $l$.

The interval length between price update can be bounded with $RTT_l$ by utilizing (4) and (5), such that

$$RTT_l \geq 2 d_l^{max} + d_l^{serv}, \quad (6)$$

where $d_l^{serv}$ is the slowest processing rate at the destination server in $l$ to process the price notification and to adjust the transmission rate in $l$. So the maximum interval length $RTT_{max}$ can be estimated as follows.

$$RTT_{max} = \max(\{RTT_l^t\}) + \epsilon, \quad \forall s \in l, \quad (7)$$

where $\epsilon$ is a positive constant error term and $RTT_l^t$ denotes $RTT_l$ over a period time of $t \to \infty$. Hence, the time for network to send price notification and receive user's response messages can be estimated by $RTT_{max}$, which also implies the price update notification interval can be estimated by $RTT_{max}$. From this point onward, $RTT_{max}$ is referred as RTT for the rest of the paper.

Next, we investigate the relationship between network price and RTT.

**Proposition 1**. RTT fluctuation corresponds to price change.

*Proof*:

*Case 1*: $x_s > x_s^{min}$.

By using Little theorem [9], we have delay function $(x_s) = \frac{\rho_s}{x_s(1-\rho_s)} = \frac{1}{x_s - x_s^{min}}$, where $\alpha_s$ is the arrival rate on $l$. Since function $d(x_s)$ is influenced by $\rho_s$, where $\rho_s = \frac{x_s^{min}}{x_s} < 1$, and the rate allocation $x_s$ is adjusted according to the network price $\lambda_s$ by solving equation (2) and (3). The changes in price $\lambda_s$ influences $\rho_s$, which affects $d(x_s)$. Thus, in this case, the fluctuation of RTT of (7) corresponds to price change.

*Case 2*: $x_s \leq x_s^{min}$.

In the occurrence of rate allocation below the required minimum bandwidth, $x_s^{min}$ is adjusted to $x_s$, such that $x_s^{min} = x_s$, because it is physically impossible to transmit more data than the available bandwidth. It is because as $\rho_s \to 1$, number of packets in the system approaches to infinity [9]. Thus, in this case, the delay function is formulated

$$d(x_s) = \frac{B_s}{x_s}, \quad (8)$$

where $B_s$ denotes the buffer size allocated for user $s$ or the maximum number of packets that the buffer can hold. Observe in (8) that the changes in $\lambda_s$ also influences $x_s$, which in turn affects delay function $d(x_s)$ because rate allocation $x_s$ is adjusted according to the network price $\lambda_s$ by solving equation (2) and (3). ∎

The Proposition 1 describes the relationship between network price and delay, such that any changes in price will impact the length of RTT. This proposition also shows that the change in RTT is proportional to the change in pricing. For this reason, pricing is employed as one of the factors to estimate RTT in LS estimation algorithm.

## IV. LEAST SQUARE ESTIMATION

In this section, we introduce LS estimation algorithm to predict RTT on the basis of the past historical information on aggregated input of network price and RTT. The section on

RTT includes a discussion on prediction on user demand for bandwidth using the same approach.

Let $h$ be the observed time interval between two iterations, so

$$h = RTT + \epsilon. \quad (9)$$

Furthermore, the price $\lambda^{(t)}$ is merely a weighted price of past *observed* interval $h^{(t)}$ over $t$ iteration, which is time shifted by appropriate update between source and network.

$$\lambda^{(t)} = h^{(t)} \, w. \quad (10)$$

Let $\hat{h}^{(t+1)}$ be the estimate time interval between two iterations. Given the past history $t$, for $t \to \infty$, observed intervals $h$ and network price $\lambda$, the network estimates the weight distribution $w$, which is used to determine $\hat{h}^{(t+1)}$, especially when the price notification packet that corresponds to the longest path is lost. The estimated time interval is defined as follows.

$$\hat{h}^{(t+1)} = \frac{\lambda^{(t+1)}}{w}. \quad (11)$$

The model above is formulated into an LS problem by introducing an error term $\varepsilon^{(t)}$ so that

$$\lambda^{(t)} = h^{(t)}.w + \varepsilon^{(t)}.$$

In a matrix form

$$\vec{\lambda}^{(t)} = \vec{h}^{(t)}.w + \vec{\varepsilon}^{(t)}$$

where

$$\vec{\lambda}^{(t)} = \left[\lambda^{(t)}, \lambda^{(t-1)}, \lambda^{(t-2)}, \dots, \lambda^{(1)}\right]^T,$$

$$\vec{h}^{(t)} = \left[h^{(t)}, h^{(t-1)}, h^{(t-2)}, \dots, h^{(1)}\right]^T,$$

$$\vec{\varepsilon}^{(t)} = \left[\varepsilon^{(t)}, \varepsilon^{(t-1)}, \varepsilon^{(t-2)}, \dots, \varepsilon^{(1)}\right]^T.$$

The LS estimator $w$ is based on the minimization of the scalar cost function $J(w)$, given by

$$J(w) = \frac{1}{2} (\vec{\varepsilon})^T \vec{\varepsilon}$$

$$= \frac{1}{2} \left[\vec{\lambda}^{(t)} - w.\vec{h}^{(t)}\right]^T \left[\vec{\lambda}^{(t)} - w.\vec{h}^{(t)}\right].$$

The LS estimator, $w$, satisfies the condition of first order derivative of $J(w)$

$$\frac{\partial J(w)}{\partial w} = \left[\left(\vec{h}^{(t)}\right)^T \vec{h}^{(t)}\right] w - \vec{\lambda}^{(t)} \left(\vec{h}^{(t)}\right)^T = 0.$$

Or $\left[\left(\vec{h}^{(t)}\right)^T \vec{h}^{(t)}\right] w = \vec{\lambda}^{(t)} \left(\vec{h}^{(t)}\right)^T$. Since $\vec{\lambda}^{(t)}$ and $\vec{h}^{(t)}$ are $m \times 1$ matrix, where

$$\vec{\lambda}^{(t)} \approx \vec{h}^{(t)}.w$$

$$\begin{bmatrix} \lambda^{(1)} \\ \lambda^{(2)} \\ \vdots \\ \lambda^{(t)} \end{bmatrix} \approx \begin{bmatrix} h^{(1)} \\ h^{(2)} \\ \vdots \\ h^{(t)} \end{bmatrix} [w],$$

the solution for $w$ is obtained as follows.

$$w \approx \left[\left(\vec{h}^{(t)}\right)^T \vec{h}^{(t)}\right]^{-1} \vec{\lambda}^{(t)} \left(\vec{h}^{(t)}\right)^T$$

$$= \frac{\sum_{t=0}^{m} h^{(t)} \lambda^{(t)}}{\sum_{t=0}^{m} (h^{(t)})^2}. \quad (12)$$

Eq. (10) indicates that network does not require a large memory space to store information. Instead of allocating space for each $h^{(t)}$ and $\lambda^{(t)}$, for $t \to \infty$, network only needs a space to store each value of $\sum_{t=0}^{m} h^{(t)} \lambda^{(t)}$, $\sum_{t=0}^{m} (h^{(t)})^2$, and $w_s$, which is $O(1)$ of memory space for each value. When new information becomes available, network simply aggregates the new information to (12). Next, we evaluate the effectiveness of LS estimation is. From this point onward and the rest of the paper, $s$ refers as user with the farthest distance from where the price notification originates.

**Proposition 2.** $w_s$ converges as $t \to \infty$.

*Proof:* Assuming there exists an optimal solution for the dual problem $D$, such that price $\lambda_s^{(t)}$ and rate allocation $x_s^{(t)}$ converges at time $t$, as $t \to \infty$. With this assumption, $\rho_s^{(t)}$ also converges, where $\rho_s^{(t)} = \frac{x_l^{min}}{x_s^{(t)}}$, for $l \in r(s)$. Since delay function $d_l^{info}$ in eq. (4) is influenced by $x_s^{(t)}$ through $\rho_s^{(t)}$, then $d_s^{info}$ converges, which implies $d_l^{max}$ in eq. (5) also converges. Since time interval $h^{(t)}$ is obtained by solving (7) and (9), then $h^{(t)}$ also converges if and only if $d_l^{max}$ converges. Thus, given the relationship between $h_s^{(t)}$ and $\lambda_s^{(t)}$ in (11), $w_s$ also converges as $t \to \infty$. ∎

Proposition 2 implies that as $t \to \infty$, if there exists a solution to the dual problem $D$, then the predicted outcomes should also converge to a value, which is a similar behavior to subgradient algorithm. Additionally, in order to save memory space, instead of estimating the RTT of every user in $l$, network only needs to keep track the longest RTT of every link, which results in $O(L)$ memory space.

The gap between the estimate time interval $\hat{h}_s^t$ and the actual time interval $h_s^t$ can be summarized as follows.

$$\Delta h_s^{(t)} = \left|h_s^{(t)} - \hat{h}_s^{(t)}\right|.$$

Subsequently, given $h_s^i$ and $\hat{h}_s^i$, for $0 \le i \le t$, $\Delta h_s^{t+1}$ can be minimized, such that $h_s^{(t+1)} \approx \left|\Delta h_s^{(t+1)} - \hat{h}_s^{(t+1)}\right|$. Let

$$\Delta \bar{h}_s^{(t)} = \frac{1}{t} \sum_{t=0}^{\infty} \Delta h_s^{(t)}$$

$$= \frac{1}{t} \sum_{t=0}^{\infty} \left|h_s^{(t)} - \hat{h}_s^{(t)}\right|.$$

**Corollary 1.** $\hat{h}_s^{(t+1)}$ converges as $t \to \infty$.

*Proof:* As shown in Proposition 2 that $w_s$ converges if there exists an optimal solution for the dual problem, such that $\lambda_s^{(t)}$ converges, as $t \to \infty$. Since $\hat{h}_s^{(t+1)}$ is obtained by solving (11), so $\hat{h}_s^{(t+1)}$ also converges as $t \to \infty$. ∎

**Proposition 3.**
$$\lim_{t \to \infty} \frac{\left|\Delta\bar{h}_s^{(t)} - \hat{h}_s^{(t+1)}\right|}{h_s^{(t+1)}} = 1$$

*Proof.* First of all, we show that $\Delta\bar{h}_s^t$ converges, as $t \to \infty$. By proposition 2 and corollary 1, if there exist an optimal solution to the dual problem $D$, then $h_s^t$ and $\hat{h}_s^t$ also converge, which also implies that the mean value $\Delta\bar{h}_s^{(t)}$ also converges. For this reason, $\left|\Delta\bar{h}_s^{(t)} - \hat{h}_s^{(t)}\right| \to h_s^{(t)}$, as $t \to \infty$. Hence, we conclude that

$$\lim_{t \to \infty} \frac{\left|\Delta\bar{h}_s^{(t)} - h_s^{(t)}\right|}{h_s^{(t)}} = \lim_{t \to \infty} \frac{\left|\Delta\bar{h}_s^{(t)} - \hat{h}_s^{(t+1)}\right|}{h_s^{(t+1)}} = 1 \, .$$

∎

Proposition 3 shows that when there is sufficient information, the gap between the estimation and the actual RTT can be predicted, which also means $\Delta\bar{h}_s^{(t)}$ can be minimized. To minimize the gap, we consider two cases: First case, when the gap $\Delta\bar{h}_s^{(t)}$ between the estimate and the actual value at time $t$ is too large, error correction to minimize $\Delta\bar{h}_s^{(t)}$ may be necessary. However, in the second case, if the gap $\Delta\bar{h}_s^{(t)}$ is too small, then it may not be necessary to perform an error correction. Observe that

$$\lim_{t \to \infty} \Delta\bar{h}_s^{(t)} = \lim_{t \to \infty} \frac{\sum_{t=0}^{\infty} \Delta h_s^{(t)}}{t} = 0,$$

Let $\varepsilon_s$ denote a positive constant decided by the network, such that. The error correction is performed only when $\Delta\bar{h}_s^{(t)} > \varepsilon_s$. The objective of error correction is

$$minimize \, |\hat{h} - \Delta h|$$

$$over \, \hat{h}, \Delta h > 0.$$

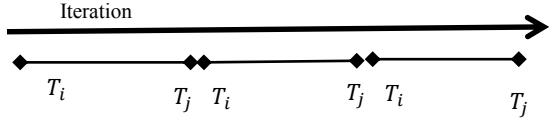

Fig. 4. Iteration Interval.

The suitable approach for *estimation error minimization* in real-time environment is when the methodology can be executed quickly without requiring heavy computing resources and a large memory space. It is because higher computation and memory space also means higher overhead cost. Let $T_i$ be the initial iteration when $\Delta h$ is observed and $T_j$ is the last iteration before the network stops observing $\Delta h$, for $i < j$, as illustrated in figure 4. The mean value of $\Delta h$ between iteration $[T_i, T_j]$ for user $s$ with the longest RTT is defined as follows.

$$\Delta h_s^{|T_j - T_i|} = \frac{\sum_{t=T_i}^{T_j} \Delta\bar{h}_s^{(t)}}{|T_j(s) - T_i(s)|} \, . \quad (13)$$

The reason for computing a new mean value of $\Delta h$ in every interval of $|T_j - T_i|$ iterations because $\Delta h_s^{|T_j - T_i|}$ is sensitive to the gradient of RTT. On the other hand, $\Delta h_s^{|T_n - T_0|}$, for $n \to \infty$, is less sensitive to the gradient. After network computes estimated $\hat{h}_s^{T_{j+1}}$ at $T_{j+1}$, $\hat{h}_s^{T_{j+1}}$ is adjusted according to this procedure.

$$\hat{h}_s^{T_{j+1}} = \begin{cases} \hat{h}_s^{T_{j+1}} - \Delta h_s^{|T_j - T_i|}, & h_s^{|T_j - T_i|} < \hat{h}_s^{|T_j - T_i|} \\ \hat{h}_s^{T_{j+1}} + \Delta h_s^{|T_j - T_i|}, & h_s^{|T_j - T_i|} > \hat{h}_s^{|T_j - T_i|}, \quad (14) \\ \hat{h}_s^{T_{j+1}} & , h_s^{|T_j - T_i|} \approx \hat{h}_s^{|T_j - T_i|} \end{cases}$$

where $h_s^{|T_j - T_i|}$ and $\hat{h}_s^{|T_j - T_i|}$ are the mean value of $h$ and $\hat{h}$ between iteration $[T_i, T_j]$. Furthermore,

$$T_j = T_i + \Gamma,$$

where $\Gamma$ denotes a positive variable decided by the network and $\Gamma \leq$ RTT. In the subsequence interval, the next $T_i = T_{j+1}$. However, if the network does not receive the packet which carries user's respond message by time $T_{j'}$, where $T_{j'} = T_i +$ RTT, then $T_j = T_{j'}$. Then, we have

$$T_j = \begin{cases} T_{j'}, & T_i + \Gamma > T_{j'} \\ T_i + \Gamma, & otherwise \end{cases}$$

In other words, the packet with user's response for bandwidth demand can be considered lost when it does not arrive after the last recorded RTT.

In addition to estimating RTT, we extend our study to estimate user demand for bandwidth by using LS estimation. This allows us to further evaluate the performance of the estimation algorithm. The problem of bandwidth estimation can be reformulated by modifying equation (10) to

$$\lambda^{(t)} = x^{(t)} \, w \, .$$

The solution for $w_x$ is obtained as follows.

$$w_x \approx \frac{\sum_{t=0}^{m} x^{(t)} \lambda^{(t)}}{\sum_{t=0}^{m} (x^{(t)})^2},$$

where $x^{(t)}$ is the observed rate allocation at time $t$. Next, the estimated rate allocation $\hat{x}^{(t+1)}$ is processed as follows.

$$\hat{x}^{(t+1)} = \frac{\lambda^{(t+1)}}{w_x}.$$

Thus, the aggregated estimated flow on link $l$ at $(t+1)$ is $\sum_{s \in l} \hat{x}_s^{(t+1)}$. Thus, this allows network to have a better picture of the mapping of user demands.

## V. DISCUSSION AND SIMULATION

In this section, we present the performance results of LS estimation in three different scenarios.

### A. Scenario One

In this scenario, we compare the estimated RTT to the actual RTT in a single link network environment with link capacity of 10 shared by 3 users. Initially, each user transmits data at 10 units per second, which results in network congestion. Thus, the network resolves the congestion by solving (2) and (3). In this scenario, the estimated and actual RTT is computed as the network resolving the congestion. Given the history of the longest RTT observed between time 0 and $t$, the network estimates RTT at $t+1$ by solving eq. (11) and eq. (12). The error is minimized by solving eq. (13). Since there is no data available at the initial stage, the network only begins to make estimation after successfully receiving the first packet with user's respond message.

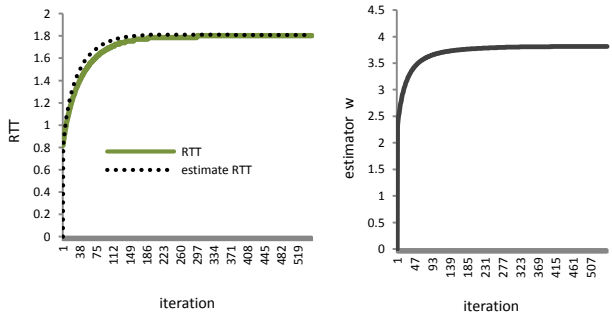

Fig. 5. RTT and estimator $w$ convergence.

The results in figure 5 illustrates that the estimator $w$ and the estimated RTT asymptotically converge. Moreover, the graph also shows that the estimated and the actual RTT behave similarly. It is because a longer historical record produces closer estimation, especially after the algorithm stabilizes. Thus, this outcome confirms our theoretical result that the algorithm with LS estimation can achieve convergence. In the following scenario, we simulate LS estimation in a larger network with different situations of failure rate.

### B. Scenario Two

Here, we present the performance of estimated RTT in various failure rates with a larger network (parking lot topology) shared by three users (user 0, 1, and 2), as depicted in figure 6. The user configuration is described in table 1. Each link has a capacity of 10 and each user initially transmits data at 10 units per second. Thus, link BC and CD become congested. Similar to scenario one, the network congestion is resolved with subgradient based algorithm. Figure 7 shows that the actual RTT converges in this setup, which also means that the subgradient based algorithm. With this algorithm convergence, we can attribute subsequence outcomes observed in this simulation to incorporation of LS estimation. Since the behaviors of three users are identical, simulation results only focus on user 0.

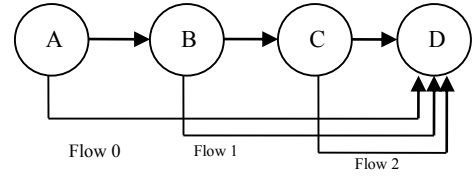

Fig. 6. Parking Lot Topology.

|  | User 0 | User 1 | User 2 |
|---|---|---|---|
| Distance from D | 1 | 2 | 3 |

Table 1. Simulation setup.

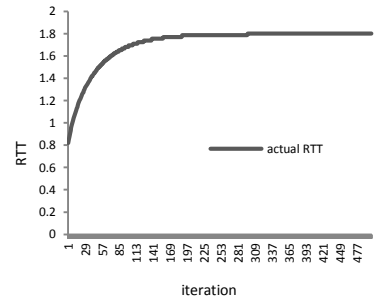

Fig. 7, The actual RTT.

In this scenario, notification packet loss is introduced at different rate: Every $50^{th}$, $40^{th}$, $30^{th}$, $20^{th}$, $10^{th}$, and $5^{th}$ iteration. During the occurrence of packet lost, network predicts the next RTT value according to previous information of the actual RTT. Furthermore, the simulator randomly decides whether packets with user's response message or price notification are dropped. For the purpose of analysis, the entire packets of the selected notification are dropped to make the inaccuracy to be more visible for analysis. Whichever notification packets are dropped, the network will not receive updates from users.

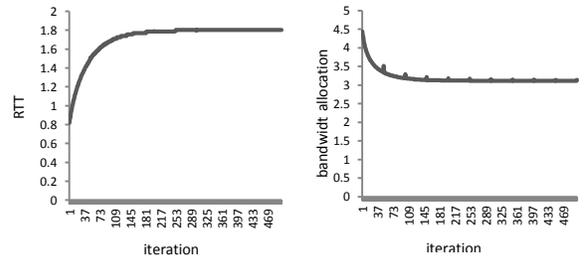

Fig. 8: Failure rate of every 50 iteration.

The simulation begins with failure rate at every $50^{th}$ iteration, where the network must predict the RTT at every $50^{th}$ iteration. Notice in figure 8, RTT with the packet loss behaves similarly to the RTT without the packet loss in figure 7. That is the RTT converges smoothly even with notification packet lost in every $50^{th}$ iteration. This means the estimation value falls on the line of the actual value. In other words, the network makes a close estimation to actual.

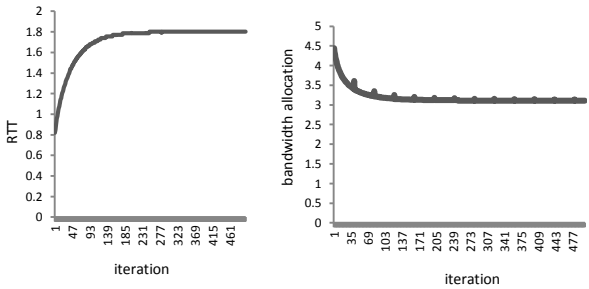
Fig.9: Failure rate of every 40 iteration.

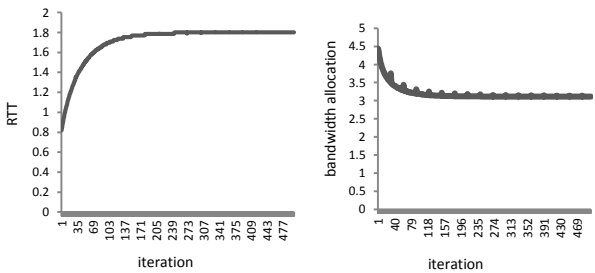
Fig. 10: Failure rate of every 30 iteration.

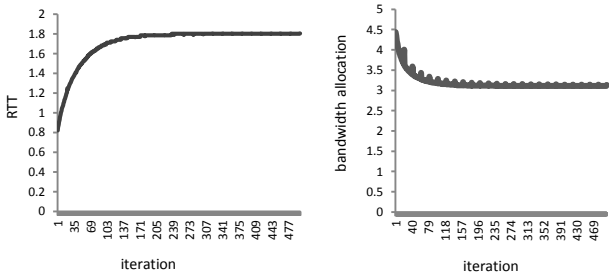
Fig. 11: Failure rate of every 20 iteration.

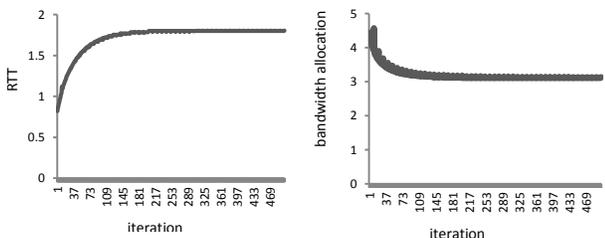
Fig. 12: Failure rate of every 10 iteration.

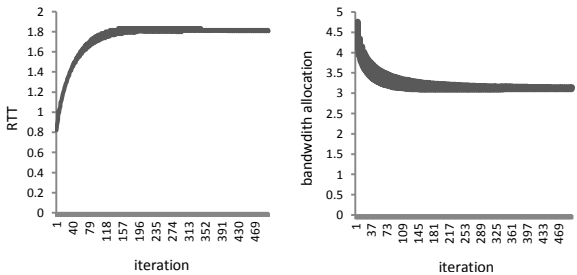
Fig. 13: Failure rate of every 5 iteration.

In the next simulation results (illustrated in figures 9 to13), the graphs show as the number of notification packets loss increases, the RTT convergence becomes less smooth and the line in graph grows thicker. The line spikes whenever there is a gap between the actual and the estimated RTT; a larger gap leads to a sharper spike, which causes the line to be thicker. Thus, the line thickness in these graphs indicates the estimation accuracy in predicting RTT when the information is lost. This phenomenon becomes more noticeable in figure 13 when the packet lost occurs in every $5^{th}$ iteration. These results demonstrate that estimation degrades as more information is missing and there is less actual data to compare the estimation. However, despite the estimation degradation, the algorithm still able to achieve converges, especially when the estimation is computed in real time setting.

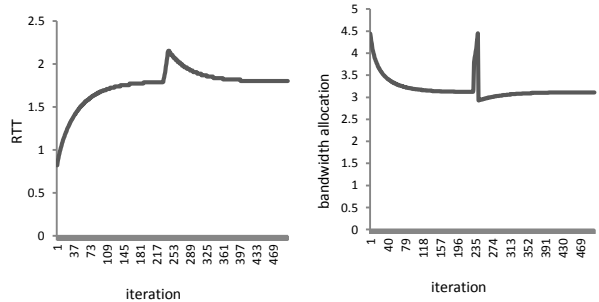
Fig. 14: Notification packet lost between iteration [230,240].

In the following simulation using the same set up as previous simulation, we present the performance result of LS estimation algorithm estimating RTT when the notification packet lost occurs sequentially. In this simulation, the information lost occurs between iteration 230 and 240. The result in figure 14 illustrates that the estimated RTT becomes inaccurate and spike grows very sharply between iteration 230-240. This shows that LS estimation does not perform well when there is sequential missing information. This is because LS estimation relies on historical information to make the prediction. When there is sequential information lost, LS estimation incorporates the previous estimated value to predict the next value. Thus, less accurate estimation leads to further inaccuracy in estimation. Based upon results from this scenario, we can conclude that LS estimation algorithm becomes less effective when there are too many packets are dropped, especially when there is loss sequential of information loss.

C. Scenario Three

To further the analysis of the performance of LS estimation, we extend our experiment to estimate user request for bandwidth allocation. In this simulation, user's demand for bandwidth is included in the ACK packet. This information is used by network to determine how much bandwidth should be allocated to each user. This simulation is designed specifically to further test the performance of LS estimation algorithm.

The result in figure 8 illustrates that the line spikes in every interval of $50^{th}$ iterations. It is because the corrected estimated value differs from the actual value. Notice that the spike size subsides as the algorithm converges. This is because after the convergence, the demand for bandwidth stabilizes and the estimation becomes more accurate. Furthermore, as the rate of packet loss increases, the bandwidth demand line grows thicker and the spikes taller, as depicted in figures 9 to 13 relative to those in figure 8. In other words, user's demand for bandwidth becomes less accurate as the spikes enlarge. The reason for the inaccuracy is more feasible for bandwidth demand because a lack of information about the users. For instance, to mimic a real condition, in this simulation, we assume that user utility of eq. (1) is not known to network. On the other hand, there is more available information to estimate RTT, for example, the information on the router processing rate and the information of the local traffic where the notification originates. Similar to the result in estimating RTT, when information lost occurs sequentially between iteration 230 and 240, the line for bandwidth demand spikes higher, as illustrated in figure 14.

**Randomized failures scenario:** There are relevant observations from the previous simulation relating to this: ($i$) The interval between failures determine a system's stability. Although the estimation value oscillates closely around the actual value, shorter interval between failures leads to higher inaccuracy. ($ii$) Continuous occurrence of failures may lead to a higher divergence from the actual value, which is expected for historical based recovery system. From these observations we derive a conclusion that in interval length between failures determines the recovery speed in randomized occurrence of failures.

In this section, we have demonstrated both the strengths and limitations of LS estimation algorithm in tackling missing information caused by network congestion. The simulation results also confirm the theoretical results discussed in the previous section that algorithm which relies on feedback mechanism can achieve convergence with LS estimation. These simulation results also show that, in most cases, the LS algorithm still delivers positive outcomes, specifically providing the environment for subgradient algorithm to achieve convergence.

## VI. DISCUSSION

Our methodology using LS Estimation features a different thinking that delivers a simple and practical solution to provide some level of fault tolerance to a real time system that relies upon feedback mechanism, especially when the failure can lead to significant revenue lost. Hence we believe that it is important to orchestrate a speedy recovery with minimal cost (i.e. memory, processors, bandwidth, etc.) without significantly affecting the system. Below, we discuss some other common concerns that might be relevant to LS estimation based recovery methodology.

**Optimality:** A potential concern with estimation based recovery approach is whether the system converges to an optimal solution. Any network failure generally distracts system from performing at an optimal level. The objective of a recovery scheme is to keep the diversion from optimal solution as minimal as possible. As we have demonstrated in our experiments our estimation oscillates around the optimal solution. However, recovery speed depends on the severity of failure and size of a system. Larger failures may lead to a slower recovery process, but this is a challenge that is shared by most recovery schemes, i.e. this concern is not unique to our proposed recovery solution. In our simulation, we demonstrate that our recovery schemes provides near optimal solution even when failures occur very frequently, as long as failure does not occur continuously over many times. At the same time, our scheme achieves linear computational and $O(1)$ memory space, which is important for real-time system. This is because decision has to be made instantaneously in the presence of failures and finding optimal solution which is typically time consuming can worsen system performance.

**Scalability:** In comparison to the Internet, the network size in cloud computing or private datacenter is much smaller. Thus, our current recovery methodology is designed with smaller real-time environment in mind, where the network (or cloud computing) providers require users to pay according to users' network usage. In large networks like the Internet, further discussion will be addressed in our future work.

## VII. CONCLUSION

In this paper, we have discussed specific challenges and implications in implementing NUM when there is information lost due to notification packet loss during excessive congestion. We propose LS Estimation algorithm to resolve the problem of information loss, and the solution asymptotically leads to algorithm convergence. The estimation carried out by LS estimation techniques minimize the squared errors between the measured and predicted RTT and bandwidth demand. The LS estimation algorithm requires a solution to a linear equation. The advantage of this approach is that LS estimation algorithm can be computed linearly, does not require high computation complexity, and only requires $O(1)$ memory space for each of information stored, which is favorable for systems that operate in real-time environment. Through the simulations, we demonstrate that our LS estimation algorithm achieves desirable results.

Additionally, this paper primarily focuses on networks that actively address information loss. It is because we assume users do not have sufficient information on network traffic intensity. Thus, user relies on network's input to adjust his/her transmission rate. For this reason, in our future work, we will explore methodologies to incorporate user behavior into the estimation algorithm to achieve a higher of accuracy in prediction when there is information loss.